\documentclass[fleqn,usenatbib]{mnras}

\usepackage[T1]{fontenc}
\usepackage{ae,aecompl}

\usepackage{graphicx}	% Including figure files
\usepackage{amsmath}	% Advanced maths commands
\usepackage{amssymb}	% Extra maths symbols
\usepackage[T1]{fontenc} 
\usepackage{academicons}
\usepackage{xcolor}
\usepackage{hyperref}
\usepackage{multirow}
\usepackage{float}
\usepackage{subcaption}
\usepackage{rotating}
\newcommand{\orcid}[1]{\href{https://orcid.org/#1}{\textcolor[HTML]{A6CE39}{\aiOrcid}}}

\def\ltsima{$\buildrel<\over\sim$}
\def\lsim{\lower.5ex\hbox{\ltsima}~}
\def\gtsima{$\buildrel>\over\sim$}
\def\gsim{\lower.5ex\hbox{\gtsima}~}

\def\muv{$M_{\rm UV}$}

\def\msolyr{M$_{\odot}$~yr$^{-1}$}
\def\msol{M$_{\odot}$}
\def\mstar{M$_{\star}$}

\def\teff{\ifmmode T_{\rm eff} \else $T_{\mathrm{eff}}$\fi}

\def\fesc{$f_{esc}$}

\def\cm2{cm$^{-2}$}

\def\ewo3{$EW_{\mathrm{[O\textsc{iii}]}}$}

\def\oii{O{\sc ii}}

\def\neiii{Ne{\sc iii}}

\def\nh{\ifmmode N_{\mathrm{HI}}\else $N_{\mathrm{HI}}$\fi}

\def\vexp{\ifmmode v_{\rm exp} \else v$_{\rm exp}$\fi}
\def\taua{\ifmmode \tau_{a}\else $\tau_{a}$\fi}

\newcommand{\jwst}{{\em JWST}}
\newcommand{\hst}{{\em HST}}
\usepackage{newtxtext,newtxmath}
\title[UNCOVER high$-z$ galaxies]{JWST UNCOVER: Discovery of $z>9$ Galaxy Candidates Behind the Lensing Cluster Abell 2744}

\author[Atek et al.]{Hakim Atek$^{1}$\thanks{E-mail: hakim.atek@iap.fr},
Iryna Chemerynska$^{1}$,
Bingjie Wang$^{2,3,4}$,
Lukas Furtak$^{5}$,
Andrea Weibel$^{6}$,
\newauthor Pascal Oesch$^{6}$,
John R. Weaver$^{7}$,
Ivo Labb{\'e}$^{8}$,
Rachel Bezanson$^{9}$, 
Pieter van Dokkum$^{10}$,
\newauthor Adi Zitrin$^{5}$,
Pratika Dayal$^{11}$,
Christina C. Williams$^{12}$,
Themiya Nannayakkara$^{8}$,
\newauthor Sedona H. Price$^{9}$,
Gabriel Brammer $^{13}$,
Andy D. Goulding$^{14}$  
Joel Leja $^{2,3,4}$,
\newauthor Danilo Marchesini$^{15}$,
Erica J. Nelson$^{16}$,
Richard Pan$^{15}$,
Katherine E. Whitaker$^{7,13}$
\\
$^{1}$Institut d'Astrophysique de Paris, CNRS, Sorbonne Universit\'e, 98bis Boulevard Arago, 75014, Paris, France\\
$^{2}$ Department of Astronomy \& Astrophysics, The Pennsylvania State University, University Park, PA 16802, USA\\
$^{3}$ Institute for Computational \& Data Sciences, The Pennsylvania State University, University Park, PA 16802, USA\\
$^{4}$ Institute for Gravitation and the Cosmos, The Pennsylvania State University, University Park, PA 16802, USA\\
$^{5}$Physics Department, Ben-Gurion University of the Negev, P.O. Box 653, Be'er-Sheva 84105, Israel\\
$^{6}$Department of Astronomy, University of Geneva, Chemin Pegasi 51, 1290 Versoix, Switzerland\\
$^{7}$Department of Astronomy, University of Massachusetts, Amherst, MA 01003, USA\\
$^{8}$Centre for Astrophysics and Supercomputing, Swinburne University of Technology, Melbourne, VIC 3122, Australia\\
$^{9}$Department of Physics and Astronomy and PITT PACC, University of Pittsburgh, Pittsburgh, PA 15260, USA\\
$^{10}$Department of Astronomy, Yale University, New Haven, CT 06511, USA\\
$^{11}$Kapteyn Astronomical Institute, University of Groningen, P.O. Box 800, 9700 AV Groningen, The Netherlands\\ 
$^{12}$NSF’s National Optical-Infrared Astronomy Research Laboratory, 950 N. Cherry Avenue, Tucson, AZ 85719, USA\\
$^{13}$ Cosmic Dawn Center (DAWN), Niels Bohr Institute, University of Copenhagen, Jagtvej 128, K{\o}benhavn N, DK-2200, Denmark\\
$^{14}$ Department of Astrophysical Sciences, Princeton University, Princeton, NJ 08544, USA\\
$^{15}$Department of Physics and Astronomy, Tufts University, 574 Boston Ave., Medford, MA 02155, USA\\
$^{16}$Department for Astrophysical and Planetary Science, University of Colorado, Boulder, CO 80309, USA
}

\date{Accepted XXX. Received YYY; in original form ZZZ}

\pubyear{2023}
\begin{document}
\label{firstpage}
\pagerange{\pageref{firstpage}--\pageref{lastpage}}
\maketitle
\begin{abstract}
 We present the results of a search for high-redshift ($z>9$) galaxy candidates in the \jwst\ UNCOVER survey, using deep NIRCam and NIRISS imaging in 7 bands over $\sim45$\,arcmin$^2$ and ancillary \hst\ observations. The NIRCam observations reach a $5-\sigma$ limiting magnitude of $\sim 29.2$ AB. The identification of high$-z$ candidates relies on a combination of a dropout selection and photometric redshifts. We find 16 candidates at $9<z<12$ and 3 candidates at $12<z<13$, eight candidates are deemed very robust. Their lensing amplification ranges from $\mu=1.2$ to 11.5. Candidates have a wide range of (lensing-corrected) luminosities and young ages, with low stellar masses ($6.8<$ log(\mstar/\msol) $<9.5$) and low star formation rates (SFR=0.2-7 \msolyr), confirming previous findings in early \jwst\ observations of $z>9$. A few galaxies at $z\sim9-10$ appear to show a clear Balmer break between the F356W and F444W/F410M bands, which helps constrain their stellar mass. We estimate blue UV continuum slopes between $\beta=-1.8$ and $-2.3$, typical for early galaxies at $z>9$ but not as extreme as the bluest recently discovered sources. We also find evidence for a rapid redshift-evolution of the mass-luminosity relation and a redshift-evolution of the UV continuum slope for a given range of intrinsic magnitude, in line with theoretical predictions. These findings suggest that deeper \jwst\ observations are needed to reach the fainter galaxy population at those early epochs, and follow-up spectroscopy will help better constrain the physical properties and star formation histories of a larger sample of galaxies.
\end{abstract}


%%%%%%%%%%%%%%%%%%%%%%%%%%%%%%%%%%%%%%%%%%%%%%%%%%

%%%%%%%%%%%%%%%%% BODY OF PAPER %%%%%%%%%%%%%%%%%%

\section{Introduction}

While the {\em Hubble Space Telescope} (\hst) and ground-based observatories have uncovered more than two thousands galaxies at redshifts greater than $z \sim 6$ \citep{atek15b,finkelstein15,bouwens21}, only a handful of galaxies were known at $z >9$ \citep{oesch18, bowler20,bagley22}. This observational frontier is mainly due to the near-infrared (NIR) wavelength coverage of \hst\, which is limited to $\lambda < 2$ \micron, whereas the rest-frame ultraviolet (UV) light of early galaxies is increasingly shifted towards longer wavelengths. With its NIRCam (Near-Infrared Camera) instrument covering the $\sim 1-5$ \micron\ domain, coupled with a significantly higher sensitivity compared to its NIR predecessors \citep{rigby22,rieke23}, JWST is poised to revolutionize our views of the early stages of galaxy formation.

In the early months of operation, several studies have reported the discovery of $z>9$ galaxy candidates in the first JWST imaging observations: the Early Release Observations \citep[ERO;][]{pontoppidan22}, Early Release Science (ERS) programs CEERS \citep{bagley22b} and GLASS \citep{treu22}. Among these early results, \citet{naidu22} reported candidates at $z\sim 12-13$, \citet{finkelstein22} a candidate at $z \sim 12$, while samples of $z \sim 9-16$ candidates have been presented in \citet{atek23, donnan23, harikane23,adams23,austin23}. Many of these galaxy candidates broke the previous distance record held by \hst\ observations \citep{oesch16}. More recently, several programs have started to spectroscopically confirm some of these high-redshift candidates \citep{roberts-borsani22,morishita22}, with the highest-redshift galaxy located at $z \sim 13$ \citep{robertson22,curtis-lake23}. At the same time, the high-redshift solution of some of these candidates have been ruled out by NIRSpec follow-up observations. For example, the highest-redshift candidate at z$\sim 16.7$ \citep{donnan23} has been confirmed to be a dusty galaxy at $z\sim 4.9$ with intense rest-frame optical emission lines \citep{arrabal23}.

However, more than their distance, the most striking aspect perhaps is their combined number density and brightness. Indeed, the inferred number density is significantly larger than theoretical predictions based on galaxy formation models, or extrapolation of lower-redshift luminosity functions \citep{bouwens22b,atek23,mason22,naidu22,donnan23}. While some of these candidates have been confirmed at $z \sim 12$ or 13 \citep[e.g,][]{curtis-lake23,arrabal23}, others turned out to be low$-z$ dusty interlopers, which always warrants caution in their interpretation. Also, a sample of red massive galaxies at $z=7-9$, reported by \citep{labbe23}, appear to have stellar masses approaching that of the present-day Milky Way, in potential tension with standard $\Lambda$-CDM models \citep{boylan-kolchin22}.

Several studies have attempted to understand these early observations and interpret these surprising results. In particular, the high number density of luminous galaxies at $z>12$ has been attributed to the decreasing amount of dust attenuation at higher redshift \citep{ferrara22}, higher star formation efficiency in early galaxies and/or non-standard initial mass function \citep[IMF;][]{ziparo22,mason22}, or even non-$\Lambda$-CDM cosmologies \citep[][]{menci22,boylan-kolchin22}. In the meantime, a larger area of deep \jwst\ surveys and spectroscopic follow-up observations are needed to confirm this claim by increasing the sample size of confirmed $z>9$ galaxies.  

During the first cycle of \jwst\ operations, our UNCOVER (Ultradeep NIRSpec and NIRCam ObserVations before the Epoch of Reionization) Treasury survey has obtained deep multi-wavelength NIRCam imaging of the lensing cluster Abell 2744 \citep{bezanson22}. UNCOVER deep NIRCam imaging consists of a mosaic in 7 filters for $\sim4-6$ hour per band, reaching a magnitude limit of $\sim 29.2$ AB. Following on the steps of the Hubble frontier fields (HFF), the program relies on the gravitational lensing boost to push beyond blank fields limits. In fact, assuming an average amplification of 5, UNCOVER is intrinsically the deepest observing program of Cycle 1. In addition, the program will obtain spectra for the intrinsically faintest distant galaxies to date with 5–20 hours of NIRSpec Prism follow-up observations. In addition to our program, NIRISS imaging of A2744 was obtained as part of ERS GLASS program, and NIRCam imaging as part of the DDT program ID 2756. We combined all these imaging data to increase the depth and the area of our survey.  

In this paper, we present the detection of $z>9$ galaxy candidates in the NIRCam and NIRISS imaging data, determine their physical properties, and compare them to theoretical predictions. Using imaging data in 15 broad-band filters, the identification of galaxy candidates is based on a combination of photometric dropout criteria and photometric redshifts derived from Spectral Energy Distribution (SED) fitting with both {\tt BEAGLE} and {\tt Eazy} codes. We describe the imaging dataset in Section \ref{sec:obs} and the sample selection in Section \ref{sec:sample}. We present our estimates of the physical parameters and their redshift-evolution in Section \ref{sec:results}. Our conclusions are given in Section \ref{sec:conclusions}. We use AB magnitudes \citep{oke83} and a standard cosmology with H${_0} =70$ km s$^{-1}$ Mpc$^{-1}$, $\Omega_{\Lambda}=0.7$, and $\Omega_m=0.3$.

%--------------------------------------------------------------------
\section{Observations} \label{sec:obs}

The UNCOVER observations are described in the survey article \citep{bezanson22}, which is accompanied by our first data release of the imaging mosaics, available at the UNCOVER webpage\footnote{\url{uncover.github}}. Here we describe briefly the content of the data and their photometric characteristics.  
 
The NIRCam data consists of short wavelength SW imaging in 3 broadband filters (F115W, F150W, F200W) and long wavelength LW imaging in 3 broadband filters (F277W, F356W, F444W) and one medium band filter (F410M). The exposure times and the resulting magnitude limits in all filters are listed in Table \ref{tab:obs}. Simultaneously to the NIRCam observations, NIRISS imaging is obtained in parallel using 5 broadband filters (F115W, F150W, F200W, F356W, and F444W). Our analysis also includes data from the GLASS survey \citep{treu22} obtained with NIRISS, which adds the F090W band in a fraction of the UNCOVER area. In addition, we incorporate NIRCam imaging from the DDT program ID 2756, which uses a similar set of filters to UNCOVER, except the F410M filter, and shorter exposure times. Using the \texttt{grizli} software (Brammer et al. in prep.), the data were then reduced and drizzled into mosaics with a common pixel scale of 0.4\arcsec\ pix$^{-1}$ and a total field of view of $\sim45$\,arcmin$^{2}$.   

The cluster core of A2744 is covered by deep \hst\ imaging from the HFF program, and a slightly wider area with shallower observations from the BUFFALO program \citep{steinhardt20}. The \hst\ observations include Advanced Camera for Surveys (ACS) imaging in three filters (F435W, F606W, F814W), and Wide-Field Camera Three (WCF3) in four filters (F105W, F125W, F140W, F160W). Furthermore, the UNCOVER NIRISS parallels overlap with deep \hst\ ACS F814W imaging in the A2744 parallel field. All these observations are drizzled to the same pixel scale and aligned to the UNCOVER images. Detailed characterization of the data are presented in \citet{weaver23}.

\begin{table}
    \centering
    \begin{tabular}{c|ccc}
        Filter &Depth & Area  \\
            &  ($5~\sigma$ AB) & (arcmin$^2$) \\
    \hline
              & HST & \\
    \hline
F435W & 29.28 & 18.54 \\
F606W &  28.86 & 36.21 \\
F814W &  28.47 & 31.26 \\
F105W &  28.14 & 20.22 \\
F125W &  28.14 & 20.08 \\
F140W &  28.79 & 5.62 \\
F160W &  28.27 & 20.15 \\
		\hline
            &JWST&            \\
\hline
% Detection & 30.00 & \\
F090W &  28.93 & 12.91 \\
F115W &  28.85 & 45.12 \\
F150W &  28.87 & 45.50 \\
F200W &  28.92 & 44.71 \\
F277W &  29.34 & 44.98 \\
F356W &  29.45 & 45.67 \\
F410M & 28.85 & 28.73 \\
F444W &29.10 & 45.11 \\
    \end{tabular}
    \caption{Limiting AB magnitudes at 5$\sigma$, quoted in 0.32\arcsec diameter apertures, correspond to the area-weighted 50$^{th}$ percentiles. Area reflects the union of the LW detection footprint with that of each band.}
    \label{tab:depths}
\end{table}

\section{High-redshift Sample Selection} \label{sec:sample}

\subsection{Photometric Catalogs} \label{sec:sep}

For our sample selection and analysis, we compared two photometric catalogs: (i) the general UNCOVER catalog published in \citet{weaver23} which has been designed to fit most of the scientific investigations covered by this dataset, (ii) a custom photometric catalog specifically tailored to the detection of high-redshift galaxies. the main differences reside in the aperture size, the deblending parameters, and the aperture corrections. 

\subsubsection{General catalog}  
 The object detection and photometry are performed in images that were previously corrected for  contamination from intra-cluster light (ICL) and bright cluster galaxies, following methods developed in \citet{shipley18}. All images are matched to the point spread function (PSF) of the longest-wavelength image in the F444W filter. The detection image consists of a co-addition of the three long-wavelength \jwst\ filters F277W, F356W, and F444W. Photometry is measured in 0.32\arcsec\ apertures using the python version of Source Extractor SEP \citep{bertin96, barbary16}. We adopted the following parameters for the SEP extraction: a detection threshold of $1.5\sigma$, a deblending threshold of 16, and deblending contrast of 3e-3. The total fluxes are estimated by applying a correction derived from elliptical Kron apertures \citep{kron80}. Details of the PSF-matching procedure, and additional photometric corrections, are described in \citet{weaver23}

\subsubsection{High$-z$ catalog} 
In addition, we produce a photometric source catalog using the {\tt SExtractor} tool \citep{bertin96} in dual mode on each available image, using the F444W as the detection image. We adopt a detection threshold of 0.9 (relative to the {\tt rms} image), a minimum detection area of 6 pixels, and a deblending threshold of 3e-4. We measured individual fluxes in 0.24\arcsec\ circular apertures in each filter. The total fluxes were obtained by using a scaling factor derived from the ratio of the aperture flux to the {\tt AUTO\_FLUX} obtained from {\tt SExtractor} in the F444W image. To account for the missing flux due to the PSF wings, we measured the aperture flux as a function of the aperture radius in the PSF of the F444W band. For each object, we computed the equivalent circularized radius as $r = \sqrt{a \times b}~ \times$ {\tt kron\_radius}, and divided the encircled flux by the flux fraction in the PSF-F444W for this radius. This correction typically increases all the fluxes by  $\sim 10-20$\%. 

In the end, the comparison between the two catalogs shows that the latter is better suited for high-redshift sources, particularly in deblending the small objects, and in estimating the object and background fluxes.  

\subsection{Dropout selection} \label{sec:dropout}
 
 Following the color-color criteria defined in \citet{atek23} we select $9<z<11$ galaxies that satisfy:
 
 \begin{equation}
	\begin{array}{l}
		M_{115}-M_{150}>1.0\\
		M_{115}-M_{150}>1.5+1.4(M_{150}-M_{200})\\
		M_{150}-M_{200}<0.5
	\end{array}
  \label{eq:z9}
\end{equation}

\begin{figure*}
    \centering
    \includegraphics[width=\textwidth]{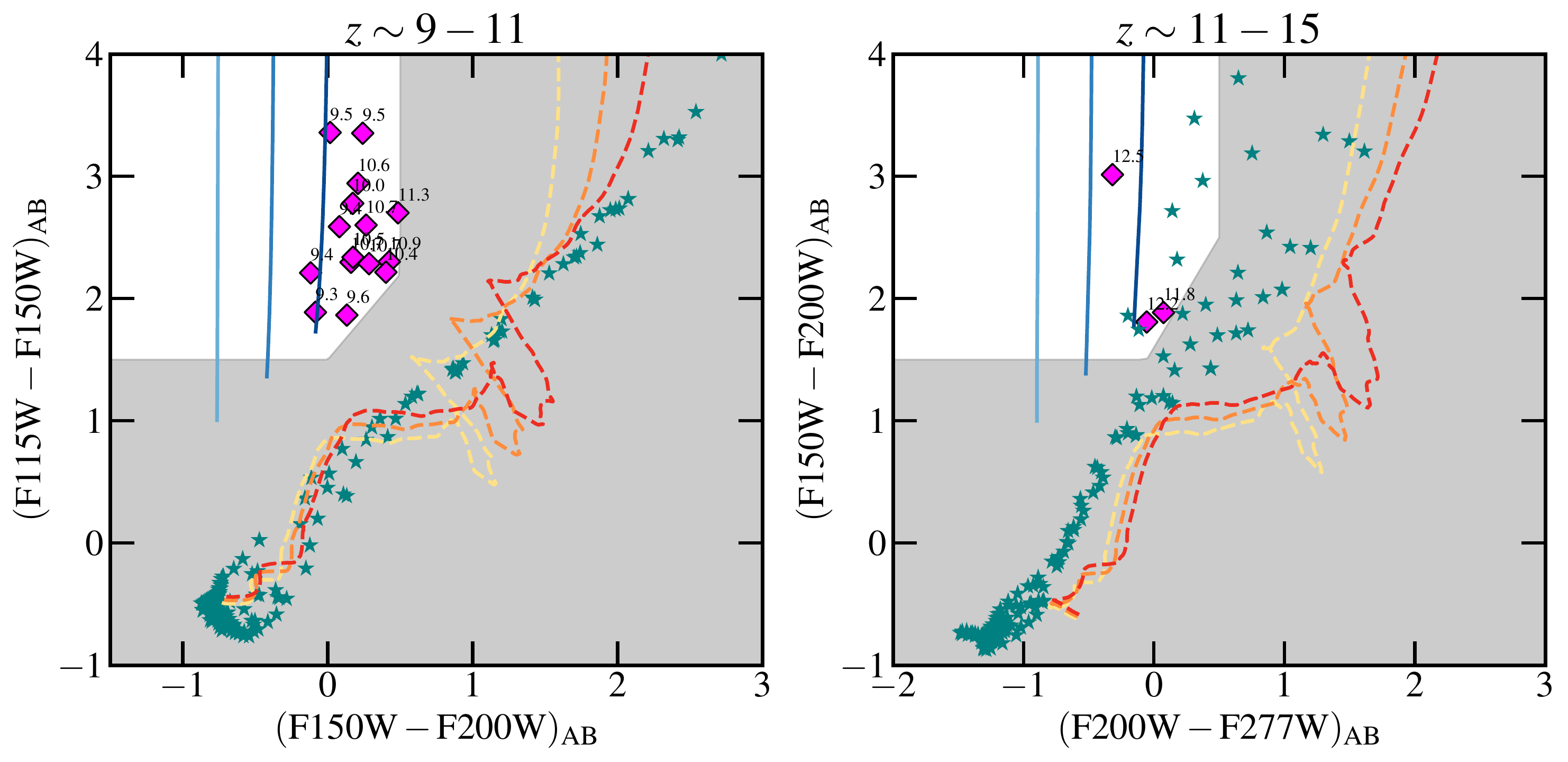}
    \caption{Color-color selection identification of high-redshift dropouts. Each panel shows the selection window (white area) defined by the criteria of equations \ref{eq:z9}, \ref{eq:z11} for the identification of candidates in the redshift range $9<z<11$ and $11<z<15$, respectively. Each candidate (magenta diamond) is marked with its associated best-fit photometric redshift. The blue-solid lines are the expected color-color space of typical starburst galaxies at $z>9$ based on galaxy templates generated using {\tt BEAGLE}. We also show the color-color tracks of quiescent galaxies (dashed lines), which represent potential low-redshift contaminants, generated from {\tt GRASIL} \citep{silva98}. We applied different dust attenuation values in the range $A_{V}=[0,0.25,1]$ illustrated by different colors (yellow to red) assuming an SMC dust law. In addition, the green stars indicate the colors of cool stars (brown dwarfs and M-class), another source of contamination, based on \citet{chabrier00} and \citet{allard01} libraries.    
    }
    \label{fig:colors}
\end{figure*}

This selection window, illustrated in Figure \ref{fig:colors}, has been designed to minimize potential contamination from low-redshift interlopers and cool stars. To determine the color-color space of these contaminants, we generated quiescent galaxy templates from the SWIRES library \citep{polletta07}, applied different dust attenuation values $A_{V}=[0,0.25,1]$ assuming an SMC dust law, and computed synthetic photometry in the set of broadband filters used in this paper. The resulting color-color tracks are shown in Figure \ref{fig:colors}. We also compute the color tracks of cold red stars and brown dwarfs using stellar templates from \citet{chabrier00} and \citet{allard01}. 

In addition to these selection criteria, we require that sources are detected in all LW filters with a minimum SNR$=5$ and that they remain undetected in F090W, when available, and all \hst\ optical bands at a $2 ~\sigma$ level. For sources that are not detected in the dropout filter, we assign a $1\sigma$ lower limit corresponding to the filter limiting magnitude to ensure a minimum continuum break of one magnitude.

For $z\sim 12-15$ candidates, we adopt the following criteria:
 
\begin{equation}
	\begin{array}{l}
		M_{150}-M_{200}>1.5\\
		M_{150}-M_{200}>1.6+1.8(M_{200}-M_{277})\\
		M_{200}-M_{277}<0.5
	\end{array}
 \label{eq:z11}
\end{equation}

Similarly, we require high-significance detection in the LW filters and that none of these candidates are detected in the bands blueward of the Lyman break, i.e. in the \hst, and \jwst\ F090W, F115W filters.

\subsection{Spectral energy distribution fitting} \label{sec:SED-fitting}

In parallel, we estimate photometric redshifts by running spectral energy distribution (SED) fitting. We apply 5\% error floor to all photometric measurements to reflect the calibration uncertainties of \jwst\ NIRCam data. To minimize the propagation of lensing uncertainties, both of the following procedures are based on the observed flux densities without correction for magnification. The derived parameters are then unlensed a posteriori. 
%% Eazy %%%%%
We first use the \texttt{Eazy} software \citep{brammer22}, over an allowed redshift range of $0.01<z<20$, adopting the {\tt CORR\_SFHZ} set of galaxy templates, which include redshift-dependent SFHs informed by the most recent results of \jwst\ observations of high$-z$ galaxies \citep{carnall23,larson23}. As shown in \citet{weaver23}, these templates perform better than the default {\tt FSPS\_FULL} library in recovering the true redshift. Improvements over the {\tt Eazy FSPS} templates have also been presented in \citet{larson22}, which include bluer UV slopes based on {\tt BPASS} \citep{stanway18} and nebular emission models.

%%% Beagle %%%%%%%%%%%%
Second, we run the {\tt Beagle} (BayEsian Analysis of GaLaxy sEds) SED-fitting software \citep{chevallard16} on the same photometric data. The procedure uses stellar population models from \citep{bc03} and nebular emission models from \citet{gutkin16}.   
%These templates assume a \citet{chabrier03} initial mass function (IMF) and IGM absorption by \citet{inoue14}.
In the first run of the SED fitting, we are mostly interested in the best redshift solution. We adopt a simple star formation history with a constant star formation rate, a uniform distribution of the age priors $\log(t_{\mathrm{age}}/\mathrm{yr})\in[7, t_{\mathrm{universe}}]$, and a fixed metallicity of $Z=0.1\,\mathrm{Z}_{\odot}$. 

In order to identify spurious objects, or sources affected by artifacts, we flag objects that meet the following criteria: objects whose segmentation apertures overlap with the edge of the detector or are next to bright stars, are affected by bad pixels, whose size is 1 pixel or less, or are likely to be stars. For the latter, we combine information from both the SExtractor stellarity parameter {\tt CLASS\_STAR} and the $\chi^{2}$ of the {\tt Eazy} SED-fitting run using a set of dwarf star templates through the {\tt fit\_phoenix\_stars} function. In addition, all the sources that pass these filters are visually inspected for potential contamination (diffraction spikes, detector artifacts, etc.)

All the candidates have best-fit photometric redshifts consistent with the dropout selection. Conversely, when first selecting candidates using photometric redshift criteria, limiting the sample to best-fit solutions with $\chi ^{2} < 30$, 31 candidates satisfy the selection. The dropout selection rejects 13 candidates. For the majority of these rejected candidates, the high$-z$ solution is due to fitting failures, where the best-fit SED does not match the photometry. Some objects do not show a clear Lyman break, or are clearly detected in the blue bands. Also, in few cases, the signal-to-noise level in the detection bands is simply below our selection threshold.

The final sample of high-redshift candidates consists of a total of 16 galaxies in the redshift range $9<z<11$ and 3 galaxies at $11<z<15$. The complete list and properties of the high$-z$ sample are reported in Tab.~\ref{tab:sample}.

\begin{table*}
\centering
\caption{Photometric and physical properties of the sample of high-redshift candidates identified through the A2744 cluster at $9<z<11$ and $12<z<15$. The photometric redshift $z_{phot}$ is derived from the {\tt Eazy} SED-fitting. The UV absolute magnitude, stellar mass and SFR are corrected by the amplification factor $\mu$, which was computed with the latest UNCOVER lensing model. Column 2 provides the quality of the candidate based on the robustness of the photometric redshift, Q=1 being the most secure candidates.}

\begin{tabular}{l|c|c|c|c|c|c|c|c|c}
ID & Q&RA & Dec & $z_{\mathrm{phot}}$ & $M_{\mathrm{UV}}$ & $\beta$ & $\log(M_{\star}/\mathrm{M}_{\odot})$ & SFR ($\mathrm{M}_{\odot} ~\mathrm{yr}^{-1})$ & $\mu$ \\\hline
\multicolumn{10}{c}{$z\sim9-11$ candidates}\\\hline 

 1870 &3 &3.648010 &-30.426616 & $9.32_{-6.95}^{+0.96}$ &-19.78 $\pm$ 0.18 &-2.09 $\pm$ 0.14 & $8.00_{-0.21}^{+0.17}$ & $1.16_{-0.45}^{+0.54}$ & $1.30_{-0.01}^{+0.01}$\\
 2065 &1 &3.617194 &-30.425536 & $9.50_{-0.08}^{+0.34}$ &-21.67 $\pm$ 0.12 &-2.03 $\pm$ 0.04 & $8.57_{--0.06}^{+0.44}$ & $4.27_{--0.58}^{+7.33}$ & $1.65_{-0.03}^{+0.02}$\\
 3148 &3 &3.646481 &-30.421615 & $9.40_{-7.14}^{+0.88}$ &-20.51 $\pm$ 0.18 &-1.92 $\pm$ 0.12 & $8.47_{-0.25}^{+0.55}$ & $3.41_{-1.51}^{+8.66}$ & $1.31_{-0.01}^{+0.02}$\\
 3160 &2 &3.591436 &-30.421663 & $10.74_{-1.45}^{+0.37}$ &-19.06 $\pm$ 0.15 &-1.79 $\pm$ 0.08 & $8.15_{--0.13}^{+0.74}$ & $1.55_{--0.58}^{+6.40}$ & $2.49_{-0.08}^{+0.09}$\\
 10619 &  1 & 3.594996 &-30.400738 & $9.69_{-0.12}^{+0.33}$ &-17.57 $\pm$ 0.13 &-2.16 $\pm$ 0.05 & $6.78_{-0.45}^{+0.35}$ & $0.68_{-0.45}^{+0.30}$ & $11.50_{-0.50}^{+0.40}$\\
17987 &3 &3.641572 &-30.382825 & $9.41_{-7.12}^{+0.60}$ &-19.39 $\pm$ 0.18 &-2.05 $\pm$ 0.14 & $7.57_{-0.45}^{+0.20}$ & $0.42_{-0.27}^{+0.25}$ & $1.31_{-0.01}^{+0.01}$\\
21623 &1 &3.567067 &-30.377869 & $10.01_{-0.26}^{+0.36}$ &-19.01 $\pm$ 0.14 &-2.30 $\pm$ 0.07 & $7.86_{-0.05}^{+0.06}$ & $0.83_{-0.09}^{+0.12}$ & $3.72_{-0.18}^{+0.14}$\\
22360 &2 &3.637111 &-30.376780 & $10.73_{-1.19}^{+0.44}$ &-19.85 $\pm$ 0.18 &-2.08 $\pm$ 0.12 & $8.33_{-0.11}^{+0.17}$ & $2.47_{-0.54}^{+1.17}$ & $1.33_{-0.01}^{+0.01}$\\
26928 &1 &3.511925 &-30.371861 & $9.47_{-0.07}^{+0.44}$ &-20.36 $\pm$ 0.12 &-1.97 $\pm$ 0.04 & $9.02_{-0.09}^{+0.26}$ & $6.54_{-0.29}^{+3.30}$ & $1.67_{-0.09}^{+0.07}$\\
31763 &3 &3.519867 &-30.366428 & $11.31_{-8.63}^{+0.20}$ &-18.89 $\pm$ 0.17 &-2.13 $\pm$ 0.11 & $7.73_{-0.13}^{+0.09}$ & $0.61_{-0.16}^{+0.14}$ & $1.92_{-0.11}^{+0.11}$\\
39074 &1 &3.590115 &-30.359743 & $10.60_{-0.31}^{+0.21}$ &-20.03 $\pm$ 0.14 &-2.21 $\pm$ 0.07 & $8.16_{-0.07}^{+0.08}$ & $1.65_{-0.24}^{+0.32}$ & $1.89_{-0.06}^{+0.05}$\\
46026 &3 &3.605690 &-30.352664 & $10.86_{-8.30}^{+0.32}$ &-19.92 $\pm$ 0.16 &-2.06 $\pm$ 0.14 & $8.31_{-0.15}^{+0.67}$ & $2.34_{-0.70}^{+8.57}$ & $1.47_{-0.03}^{+0.04}$\\
52008 &2 &3.478739 &-30.345535 & $10.37_{-1.09}^{+0.32}$ &-19.90 $\pm$ 0.14 &-2.11 $\pm$ 0.07 & $7.69_{-0.32}^{+0.15}$ & $0.56_{-0.29}^{+0.23}$ & $1.26_{-0.02}^{+0.02}$\\
73667 &1 &3.451412 &-30.321807 & $10.68_{-0.31}^{+0.21}$ &-20.55 $\pm$ 0.13 &-2.24 $\pm$ 0.05 & $8.37_{-0.01}^{+0.01}$ & $2.73_{-0.08}^{+0.07}$ & $1.17_{-0.01}^{+0.01}$\\
81198 &1 &3.451367 &-30.320717 & $10.50_{-0.66}^{+0.23}$ &-19.90 $\pm$ 0.14 &-2.33 $\pm$ 0.08 & $8.08_{-0.02}^{+0.02}$ & $1.38_{-0.06}^{+0.07}$ & $1.17_{-0.01}^{+0.01}$\\
83338 &2 &3.454706 &-30.316898 & $9.55_{-0.57}^{+0.91}$ &-19.24 $\pm$ 0.18 &-1.80 $\pm$ 0.12 & $9.29_{-0.10}^{+0.11}$ & $0.19_{-0.19}^{+0.37}$ & $1.17_{-0.01}^{+0.01}$\\ \hline

 \multicolumn{10}{c}{$z\sim11-15$ candidates}\\\hline
42329 & 3&3.513568 &-30.356804 & $11.83_{-7.93}^{+1.05}$ &-19.13 $\pm$ 0.18 &-2.05 $\pm$ 0.12 & $9.31_{--0.00}^{+0.54}$ & $0.36_{-0.35}^{+0.58}$ & $1.57_{-0.05}^{+0.06}$\\
46075 & 3&3.546722 &-30.352425 & $12.23_{-0.50}^{+1.38}$ &-19.10 $\pm$ 0.19 &-2.43 $\pm$ 0.20 & $7.72_{-0.05}^{+0.04}$ & $0.61_{-0.06}^{+0.06}$ & $1.82_{-0.05}^{+0.11}$\\
70846 & 1&3.498983 &-30.324758 & $12.50_{-0.15}^{+0.46}$ &-20.79 $\pm$ 0.12 &-2.27 $\pm$ 0.04 & $9.50_{-2.72}^{+0.82}$ & $7.06_{-6.99}^{+29.90}$ & $1.27_{-0.02}^{+0.02}$\\ \hline

\end{tabular}
\label{tab:sample}
\end{table*}

%%%%%%%%%%%%%%%%%%%%%%%%%%%%%%%%%%%%%%%%%%%%%
 \subsection{Gravitational lensing model}
 \label{sec:lensing}
 
\begin{figure}
    \centering
    \includegraphics[width=\columnwidth, keepaspectratio=true]{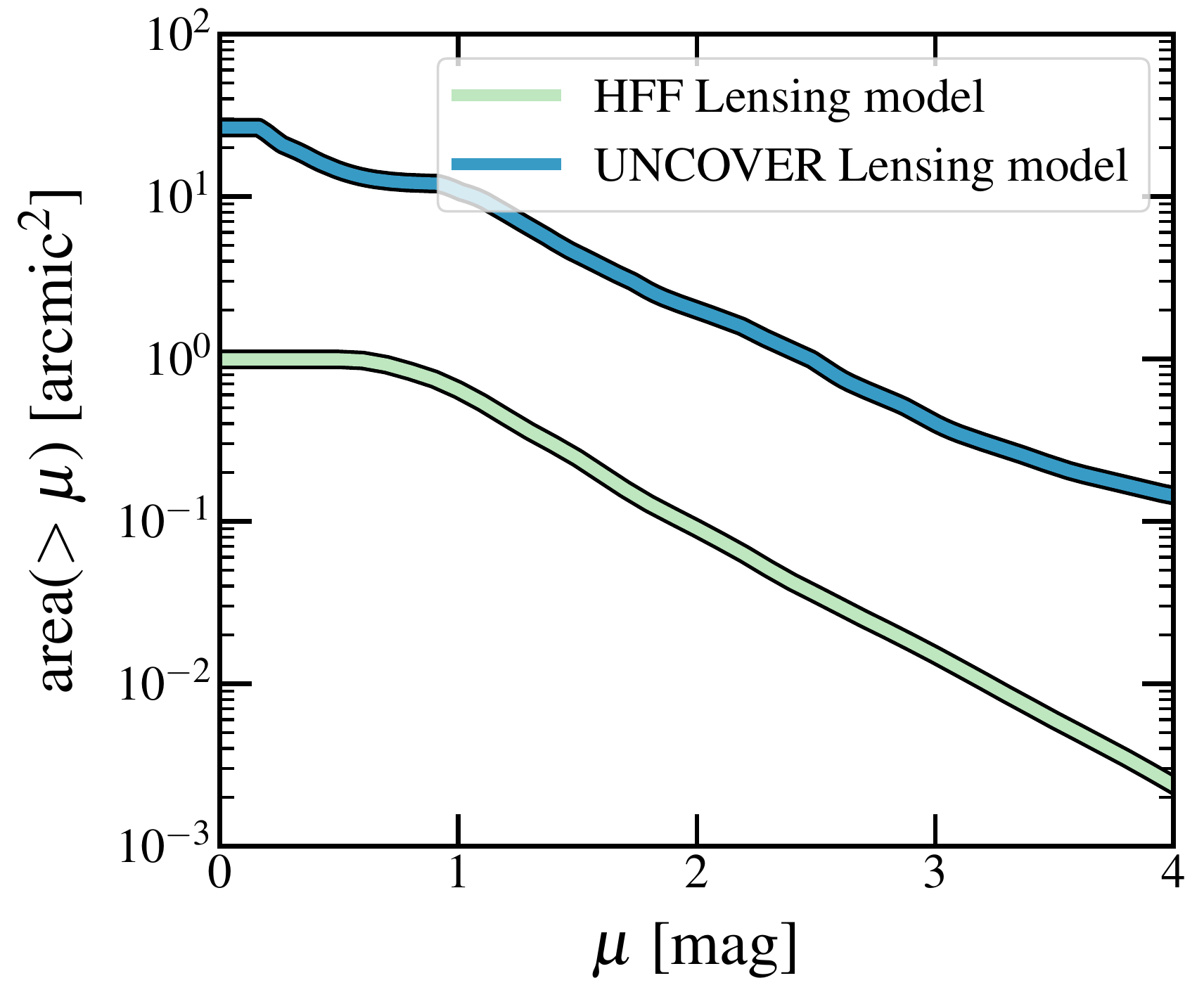}
    \caption{The cumulative surface area behind A2744 at $z=9$ (blue curve) as a function of gravitational magnification $\mu$ (cf. section~\ref{sec:lensing}), expressed in magnitudes. also shown for reference, the area as a function of magnification derived from HFF observations by the CATS team. The total unlensed survey area of UNCOVER is $\sim 35$\,arcmin$^{2}$. For reference, the observed area is about 45 arcmin$^{2}$.}
    \label{fig:area_mag}
\end{figure}

In order to compute the gravitational magnifications of the sample and the effective survey area, we use the new UNCOVER cluster mass model derived by \citet{furtak23b}. This parametric strong lensing model is based on existing and newly-discovered multiple-image systems in the deep NIRCam imaging of UNCOVER. Thanks to the wide NIRCam coverage, the total survey area with an amplification factor $\mu > 2$ is about 3.5 arcmin$^2$, which is a significant improvement over the HFF-derived area of $\sim 0.9$ arcmin$^2$. Figure \ref{fig:area_mag} shows the cumulative surface area as a function of magnification. The amplification factors shown in Table \ref{tab:sample} range between $\mu=1.2$ and $\mu=11.5$, and were derived using the photometric redshifts.

\begin{figure*}[!htpb]
    \centering
    \includegraphics[width=\textwidth]{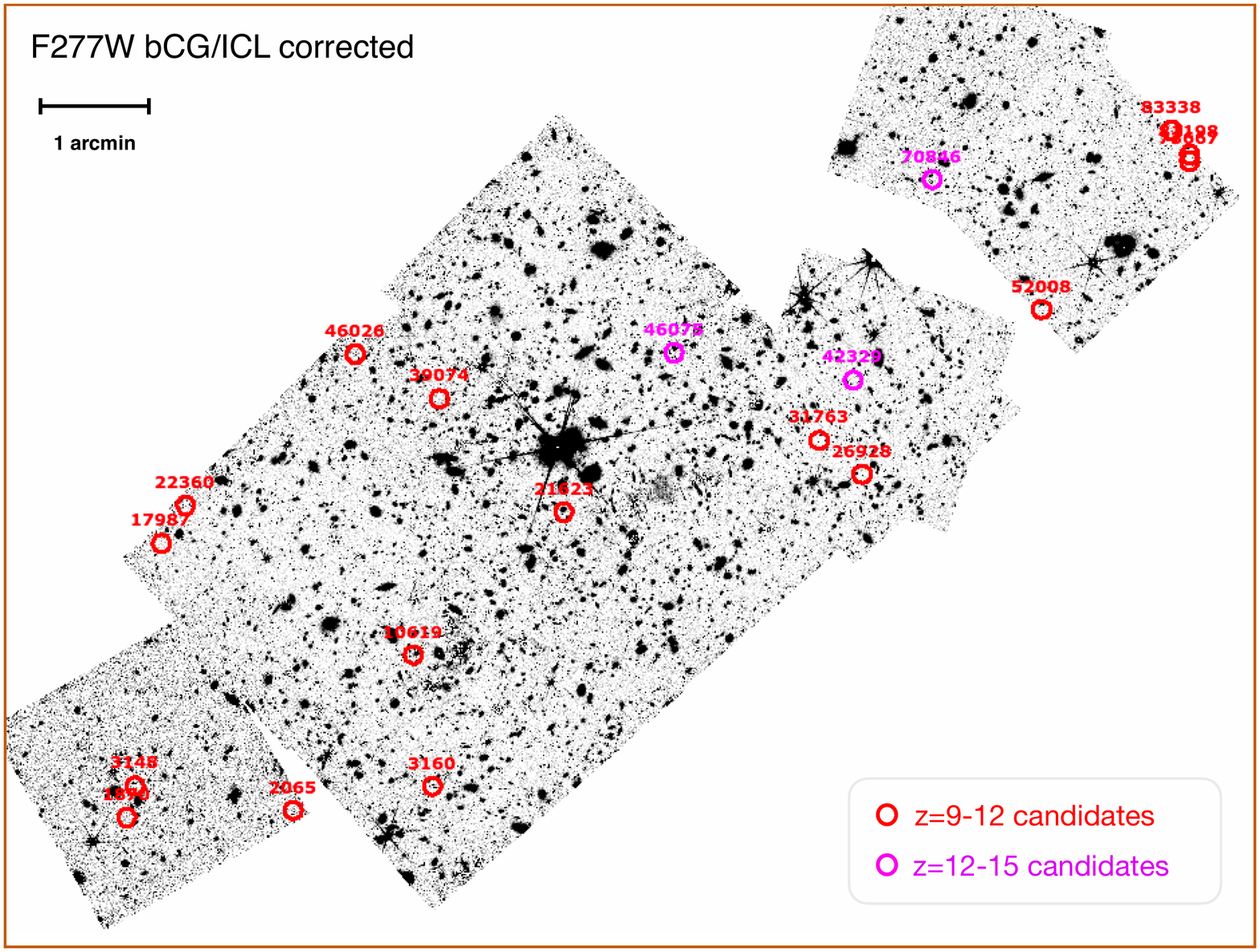}
    \caption{Coordinates of the high-redshift candidates overlaid on the F277W image, which has been corrected for bCG and ICL contamination. The footprint includes UNCOVER, GLASS, DDT observations (see text for details). The field of view of the full frame is about $12 \times 9$ arcmin.}
    \label{fig:image}
\end{figure*}
%%%%%%%%%%%%%%%%%%%%%%%%%%%%%%%%%%%%%%%%%%%%%%%%%%%%%%%%%%%%%%%%%%%%%%%%%%
\section{Results} 
\label{sec:results}

 \subsection{High-redshift candidates}

\begin{figure*}
    \centering
    \includegraphics[width=\columnwidth]{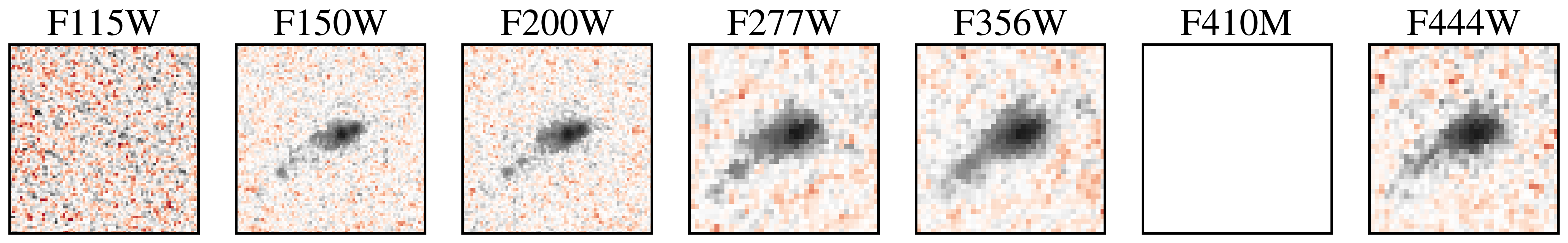}\hspace{0.5cm} 
    \includegraphics[width=\columnwidth]{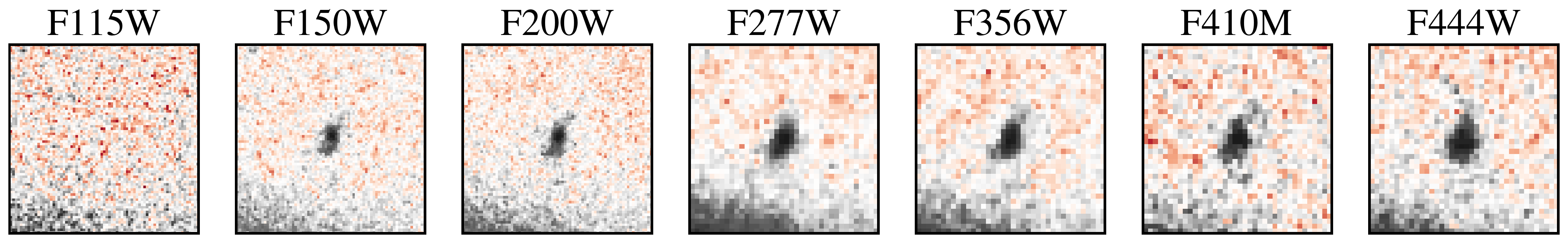}\\
    \includegraphics[width=\columnwidth]{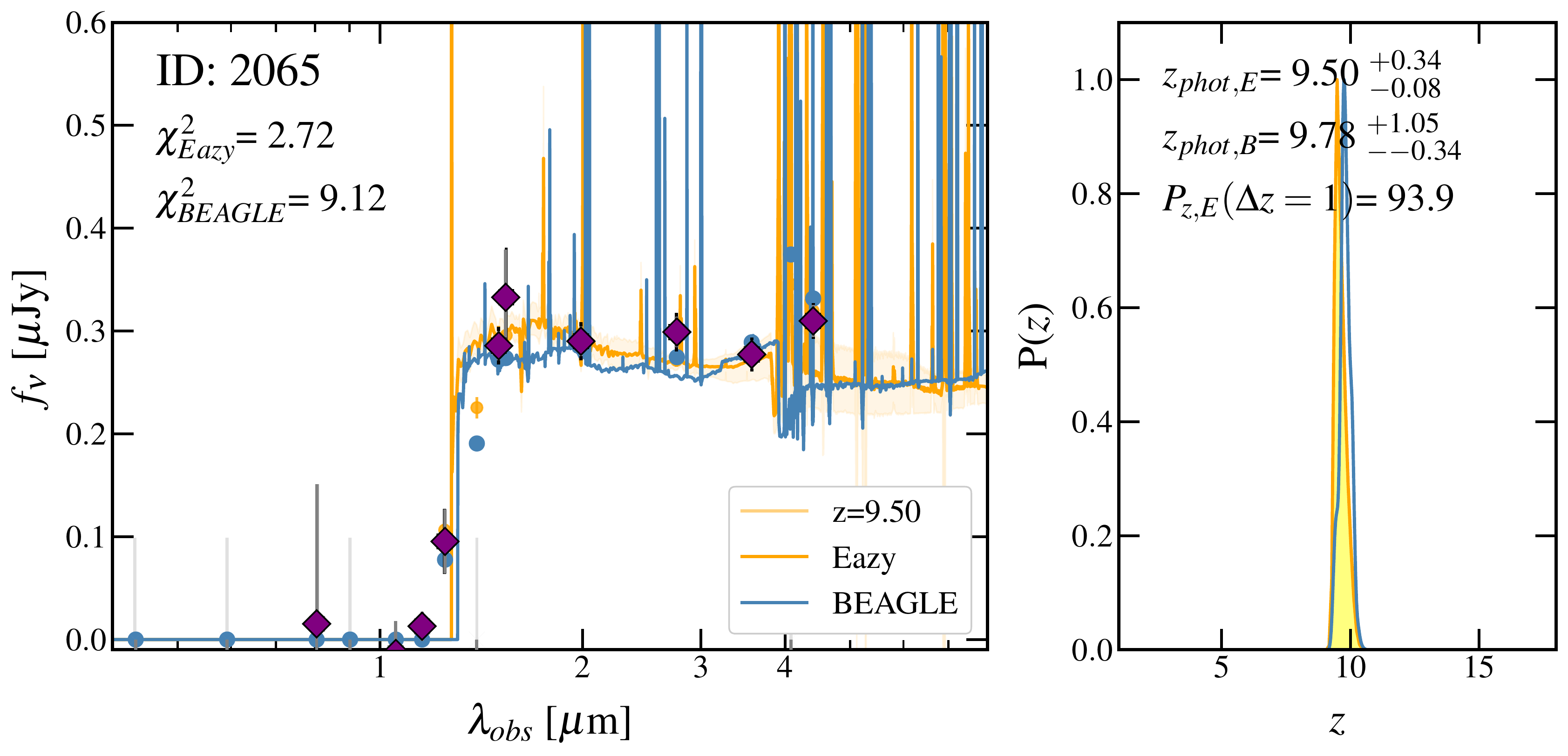}\hspace{0.5cm} 
    \includegraphics[width=\columnwidth]{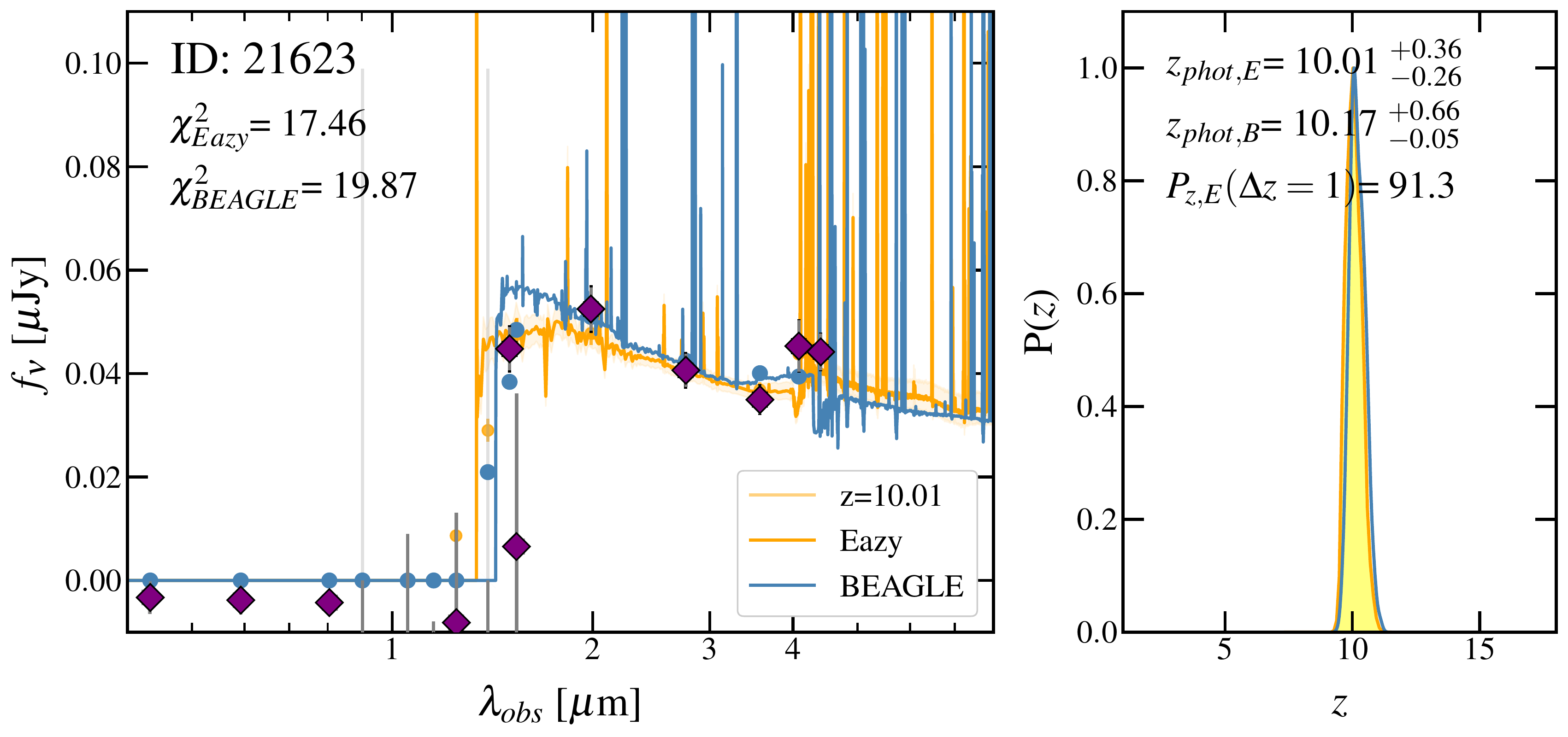}\\
    \vspace{1cm}
        \includegraphics[width=\columnwidth]{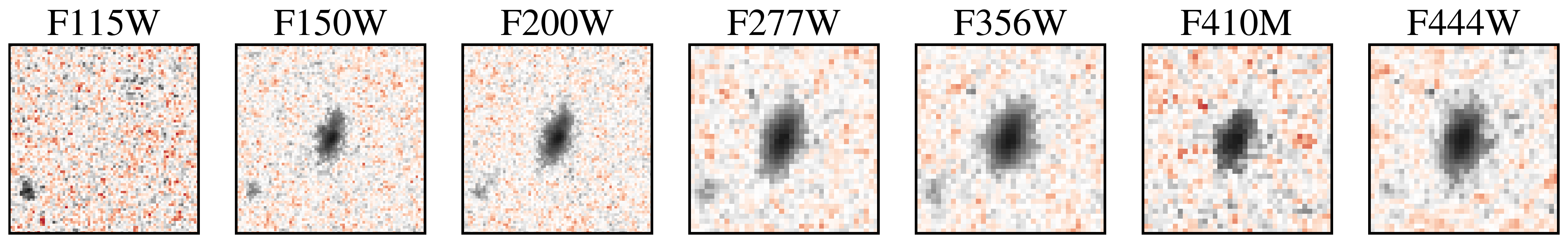}\hspace{0.5cm} 
    \includegraphics[width=\columnwidth]{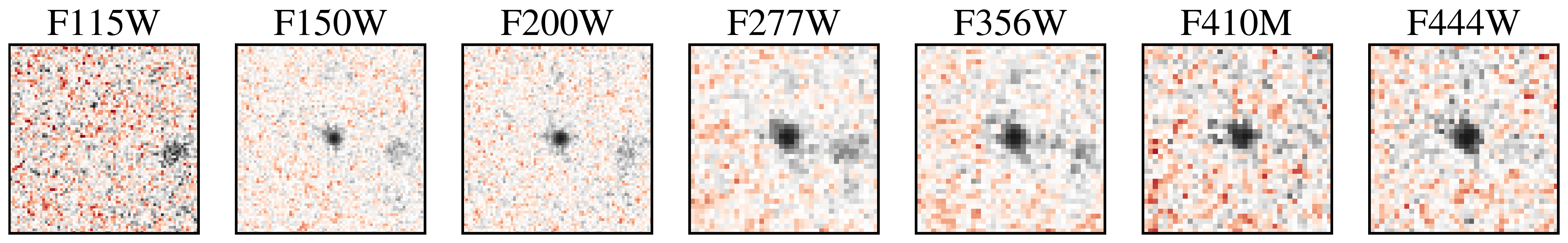}\\
    \includegraphics[width=\columnwidth]{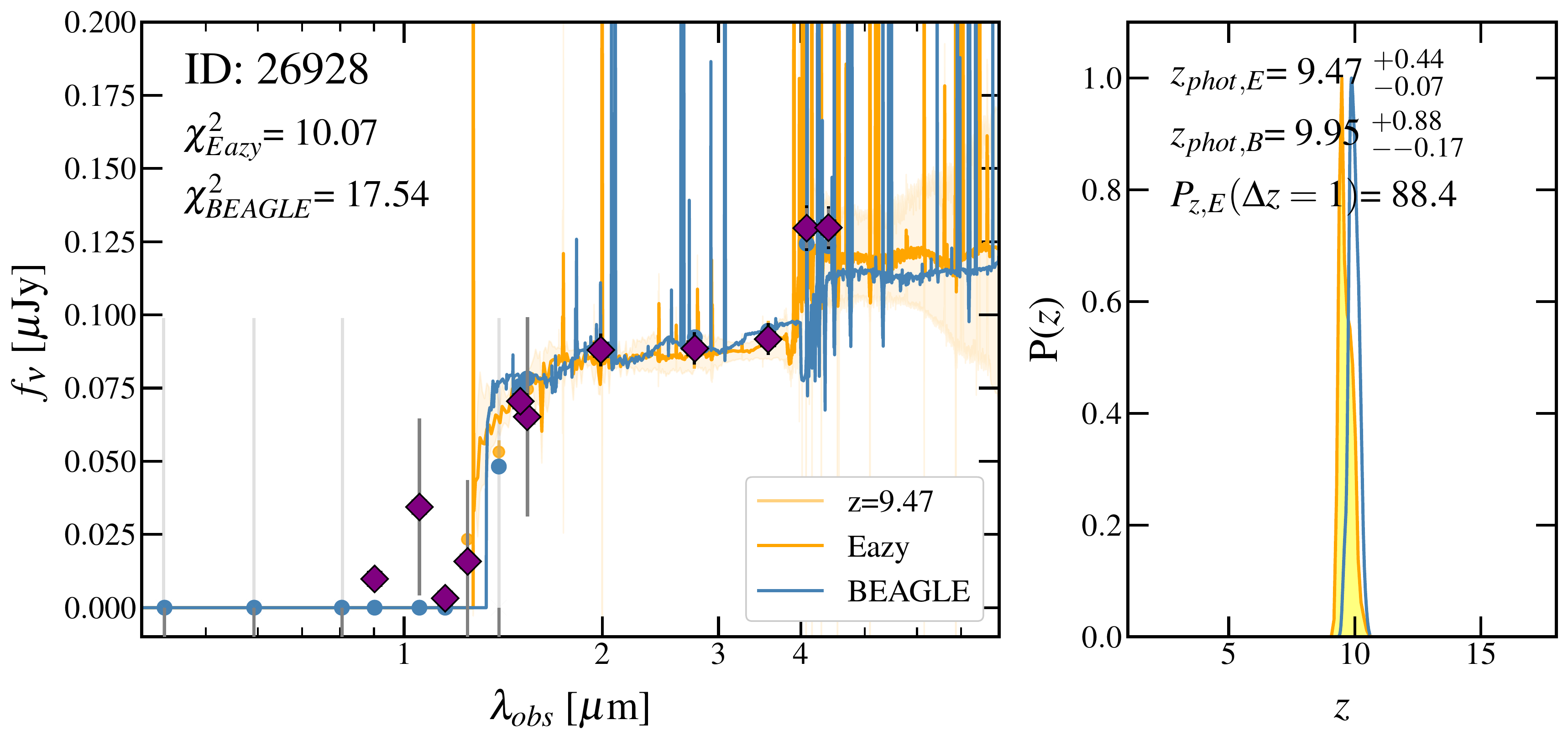} \hspace{0.5cm} 
    \includegraphics[width=\columnwidth]{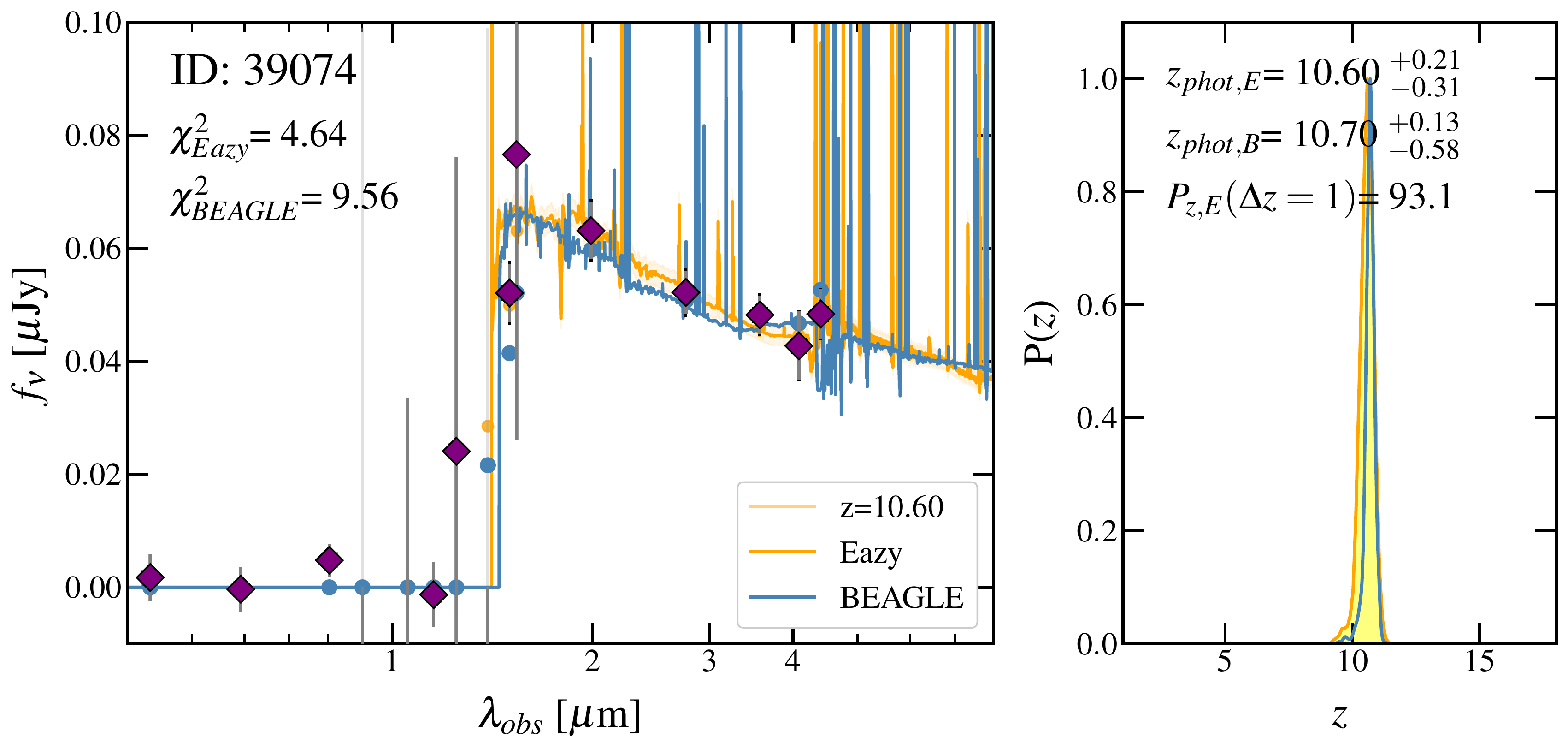}
    \caption{Imaging data and best-fit solutions for 4 of the high-redshift candidates in the range $9<z<11$. The top row of each panel shows image cutouts in the 7 \jwst\ filters. The bottom panel shows the best-fit SEDs with {\tt Eazy} (orange curve) and {\tt BEAGLE} (blue curve) together with object ID and the best fit $\chi^{2}$ from both codes. The purple diamonds represent the observed photometric data points (and their associated 1$\sigma$ uncertainties) measured in \hst\ and \jwst\ images. The orange and blue circles represent the best fit magnitudes in their respective bands for {\tt Eazy} and {\tt BEAGLE} solutions, respectively. We also show the probability distribution function (PDF) of the photometric redshift solutions (for both codes) on the right, together with the best-ft $z_{phot}$ and the total probability enclosed in a redshift width of $\Delta z =1$ around the {\tt Eazy} best-fit solution.}
    \label{fig:fits9}
\end{figure*}

We find 16 candidates in the range $9<z<11$, among which 8 objects have best-fit photometric redshifts above $z=10$. Among the total sample, we identify 7 robust candidates that have a narrow high$-z$ solution with no secondary low$-z$ solution. These candidates have about 70-90\% of their total probability enclosed within $\Delta z=1$ around the best-fit solution. Examples of these high-confidence candidates are shown in Figure \ref{fig:fits9}. These are ranked as the best candidates in the sample and are assigned a quality flag Q=1 in Table \ref{tab:sample}. The lower-quality Q=2 sample includes galaxies that show a significant secondary solution, with a total probability within 50-70\% around their high$-z$ peak. Four galaxies lie in this category. Finally, 5 sources belong to the Q=3 category because of disagreements between their {\tt Eazy} and {\tt BEAGLE} best-fit solutions or a total probability of less than 50\% around their high$-z$ peak. \citet{castellano22} recently identified 7 candidates at $z \sim 10$ in the A2744 region. We recover 6 of their candidates in our sample. Five of these overlapping sources are in the Q=1 category. One of their source, GHZ9, is ranked in the Q=2 of our sample. Source GHZ4 in their catalog is not included in our selection because it has SNR=4.8, which is slightly below our color-color selection criteria. In table \ref{tab:comparison} we compare the photometric redshifts of the sources that have been identified in both studies. Overall, there is a good agreement between these independent determinations, whose differences remain within the $1\sigma$ uncertainties, in most cases. 

\begin{figure}
    \centering
    \includegraphics[width=\columnwidth]{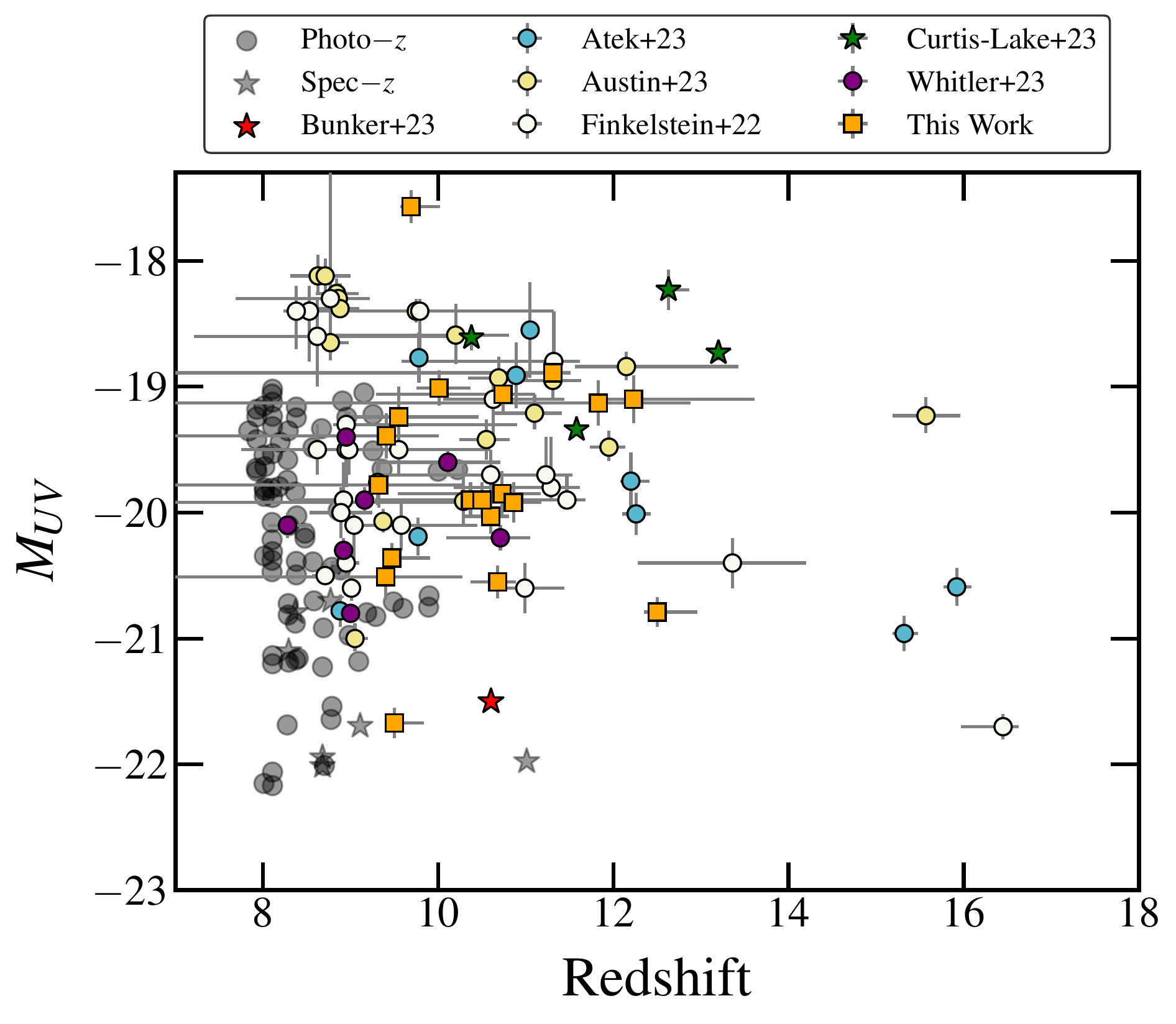}
    \caption{Absolute UV magnitude as a function of redshift for the present sample compared to literature results. The gray circles (stars) represent a compilation of known galaxies with photometric (spectroscopic) redshifts at $z >8$. The rest of the circles represents photometrically-selected galaxies from recent \jwst\ observations \citep{finkelstein22b,atek23,austin23,whitler23}. the colored stars show galaxies with spectroscopic confirmations by NIRSpec observations \citep{curtis-lake23,bunker23}.   
    }
    \label{fig:muv_z}
\end{figure}

We also identify three candidates in the redshift selection range $12<z<15$. Among these, one candidate is classified as robust Q=1, while the remaining two candidates have a Q=3 score according to the criteria defined earlier. Overall, these candidates are among the highest-redshift candidates identified in recent \jwst\ observations, as can be seen in Figure \ref{fig:muv_z}. This sample spans a large dynamic range in intrinsic luminosity from \muv=-17.6 to -21.7 mag.  

We examined the recent \jwst\ spectroscopic observations of A2744 as part of the GLASS program, utilizing both NIRSpec and NIRISS grism observations.
\begin{itemize}
    \item Candidate ID 2065 in our sample has a spectroscopic confirmation at $z=9.3$ as reported by \citet{boyett23}, in good agreement with the estimated photometric redshift of $z=9.50_{-0.08}^{+0.34}$. The NIRSpec observations of this spatially-resolved galaxy show prominent emission lines of O, Ne, and H, as well as a clear Lyman break. By combining the photometric and spectroscopic data, the best-fit SED provides a stellar mass of log(\mstar/\msol $\sim$ 9.17).    

    \item Candidate ID 10619 is one of the three multiple images of a candidate galaxy previously identified in the HFF data by \citep{zitrin14}. It has the highest magnification ($\mu \sim 11.5$). It has been spectroscopically confirmed at $z=9.76$ with NIRSpec prism spectroscopy by \citet{roberts-borsani22}. This value is in good agreement with our photometric redshift estimate of $z=9.69_{-0.12}^{+0.33}$.  
    
\end{itemize}

\begin{table}
    \centering
    \begin{tabular}{c|c|c|c|c} 
    \hline 
          ID       & GLASS ID  & $z_{phot}$({\tt Eazy})      &$z_{phot}$({\tt BEAGLE})& GLASS $z_{phot}$ \\ \hline
          2065     & DHZ1      &  $9.50_{-0.08}^{+0.34}$     &$9.78_{-0.34}^{+1.05}$  &  9.45     \\  
          
          21623     & UHZ1      &  $10.01_{-0.26}^{+0.36}$   &$10.17_{-0.05}^{+0.77}$  &  10.32     \\
          26928     & GHZ1      &  $9.47_{-0.07}^{+0.44}$    &$9.95_{-0.17}^{+0.88}$  &  10.47     \\
          52008     & GHZ9      &  $10.37_{-1.09}^{+0.32}$   &$9.47_{-0.35}^{+0.36}$  &  9.35     \\
          81198     & GHZ7      &  $10.50_{-0.66}^{+0.23}$   &$10.17_{-0.05}^{+0.66}$  &  10.62     \\
          73667     & GHZ8      &  $10.68_{-0.31}^{+0.21}$   &$10.63_{-0.51}^{+0.20}$  &  10.85     \\ 
          \hline
          
    \end{tabular}
    \caption{Comparison between the photometric redshifts derived in our analysis with {\tt Eazy} and {\tt BEAGLE} with the results of \citet{castellano22} for common objects in the two samples. 
    }
    \label{tab:comparison}
\end{table}

\begin{figure*}
    \centering
        \includegraphics[width=0.9\textwidth]{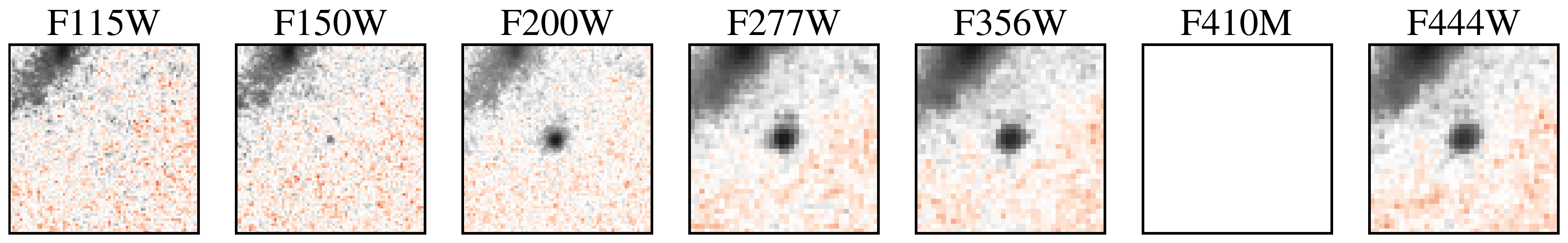} \\
    \includegraphics[width=0.9\textwidth]{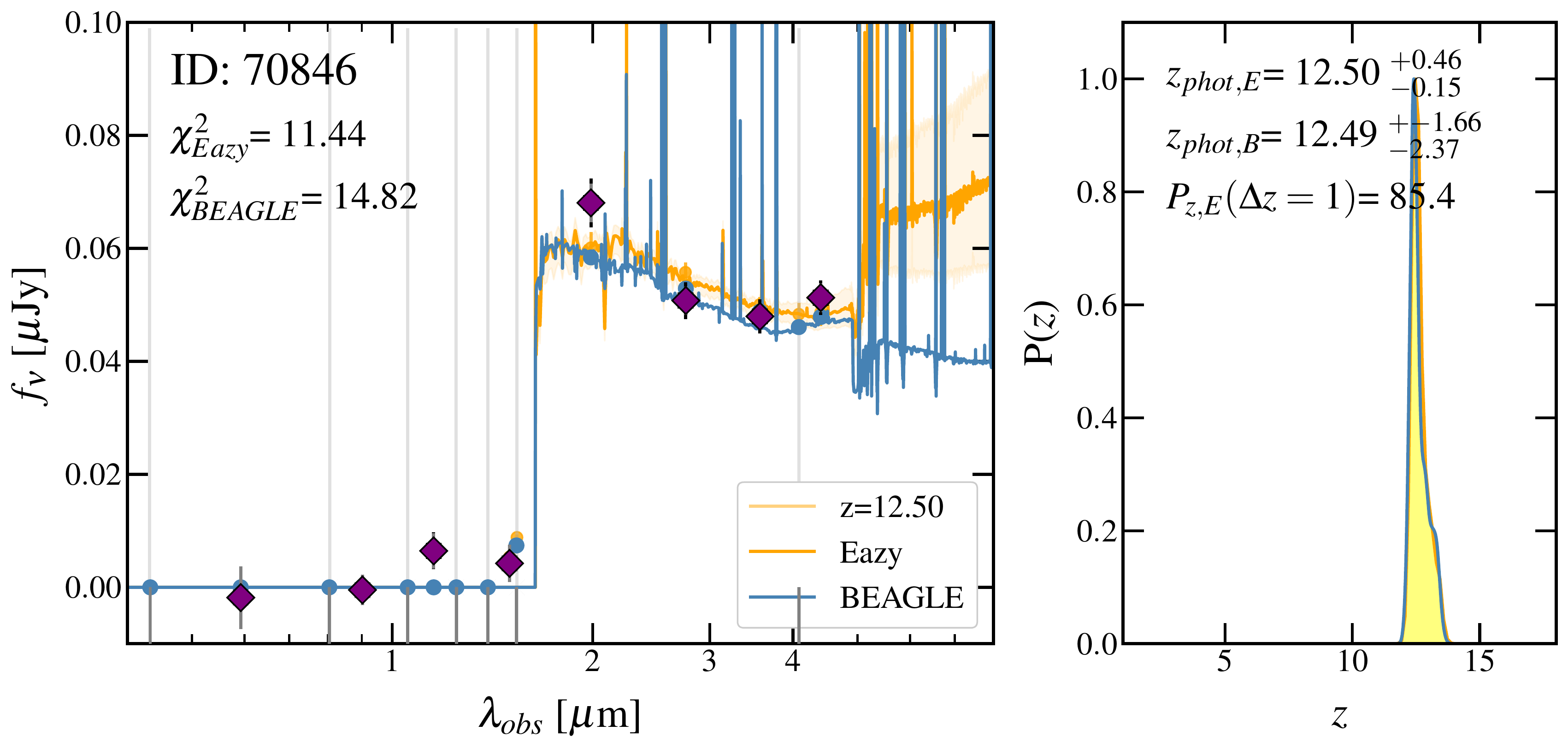} 
    \caption{same as figure \ref{fig:fits9} but for one of the candidates in the $12<z<15$ range.}
    \label{fig:fits11}
\end{figure*}

%%%%%%%%%%%%%%%%%%%%%%%%%%%%%%%%%%%%%%%%%%%%%%%%%%%%%%%%%%%%%%%%%%%%%%%
\subsection{Physical properties}

In addition to computing photometric redshifts, we perform a second SED fitting run with {\tt BEAGLE} to refine our estimates of physical parameters, this time using a Gaussian prior for the redshift based on the first-run photo$-z$ solution. We adopt this time a more flexible SFH using a delayed exponential form SFR $\propto$ t~exp($-t/\tau$), and a potential SF burst episode in the last 10 Myr. We use a constant metallicity of $Z=0.1 Z_{\odot}$, which has been shown to have little effect on the photometry of high-redshift galaxies \citep{furtak21}. We also assume an SMC extinction law, which is more appropriate for high-redshift galaxies \citep{capak15,reddy18a}. We limit the fit to 4 physical parameters using the following priors:
\begin{itemize}
    \item Stellar mass with a log-uniform distribution prior in the range log(\mstar/\msol) $\in$ [6-10]
    \item SFR averaged over the last 10 Myr with a log-uniform distribution prior log($\psi$ /\msolyr) $\in$ [-4,4]
    \item Maximum stellar age for $t$=$\tau$, with a log-uniform prior log($t_{age}$/yr) $\in$ [6, $t_{universe}$], where $t_{universe}$ is the age of the universe at the redshift of the galaxy. 
    \item Dust attenuation as traced by the optical depth measured in the V band  with a uniform prior $\tau_{V} \in$ [0, 0.5]. The prior distribution is based on UV continuum slope values measured in Section \ref{sec:beta}.   
\end{itemize}

Most of the candidates have relatively low stellar masses in the range Log(\mstar/\msol)$=6.8-9.5$. The data cover the redshifted Balmer break for most galaxies, which helps  constrain the older stellar population. Indeed, the best constraints on the stellar mass, but also the age of the stellar population, are obtained for galaxies that show an excess in the F444W band indicative of Balmer break. For example, among the robust candidates, IDs 21623, 26928 and 83338, show a significant Balmer break (Fig. \ref{fig:fits9}) and small uncertainties on their derived stellar masses. It is also interesting to note that in comparison to {\tt Eazy},{\tt BEAGLE} attribute more broadband flux to strong emission lines, which result in lower stellar masses. Strong [\oii]$\lambda\lambda 3726,3729$ and [\neiii]$\lambda 3869$ emission lines can enhance both F410M and F444W fluxes and mimic a Balmer break. Such strong lines have been observed at $z\sim 10.6$ for instance in the NIRSpec spectrum of GN-z11 \citep{bunker23}. This explains the difference observed between the {\tt Eazy} and {\tt BEAGLE} fits for a few objects, such as ID 21623. Candidates also show small SFR values, and young stellar ages between 10 and 100 Myr, confirming previous \jwst\ results at similar redshifts \citep{furtak21,topping22,austin23,whitler23}. Taken at face value and considering a constant star formation, their stellar mass would imply older ages. It is clear that the current SFR derived from our SED fitting is not representative of the entire star formation history of these candidates. Intermittent episodes of intense star formation, or simply a higher SFR, likely occurred in the past in order to build up the estimated stellar mass so quickly.  

%%%%%%%%%%%%%%%%%%%%%%%%%%%%%%%%%%%%%%%%%%%%%%
\subsubsection{Mass-Luminosity relation}

\begin{figure}
    \centering
    \includegraphics[width=\columnwidth]{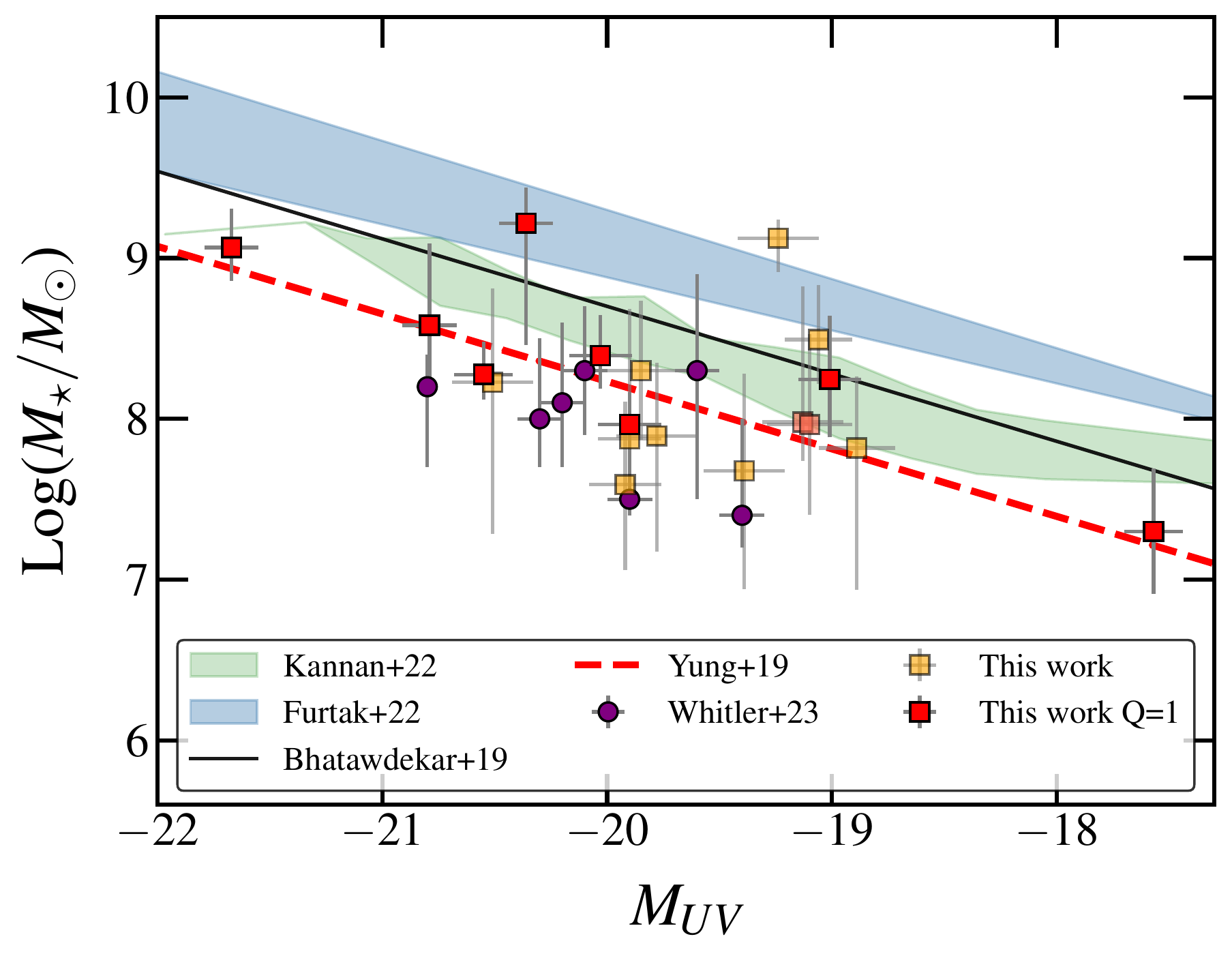}
    \caption{The stellar mass-luminosity relation for high-redshift galaxies. The sample of the present study is represented with orange squares. Best-fir relations derived from observational constraints at $z\sim6$ are indicated with the blue-shaded region \citep{furtak21}, while results at $z\sim9$ are indicated with a black line \citep{bhatawdekar19}. Theoretical predictions from semi-analytical models at $z=10$ \citep{yung19} and hydrodynamical simualtions at $z =9$ are plotted with a red-dashed line and green-shaded region, respectively.  
    }
    \label{fig:mass_muv}
\end{figure}

In addition to the stellar mass derived from the SED fitting, we computed the absolute rest-frame UV magnitude by combining the observed magnitude in F200W band and the photometric redshift (column 6 of Table \ref{tab:sample}). The \mstar-\muv\ relation provides insights into the stellar mass build up of galaxies and its evolution with redshift. Figure \ref{fig:mass_muv} shows the \mstar-\muv\ best-fit relation determined at $z\sim6$ \citep[blue line;][]{furtak21} and at $z\sim9$ \citep[black line;][]{bhatawdekar19} using \hst\ and {\em Spitzer} observations. The redshift-evolution of this relation can already be observed. Properties of the present sample are shown with squares. The red squares indicate the robust subset of candidates (cf. Table \ref{tab:sample}). Our results are consistent with a redshift evolution, where galaxies are on average below the established relations at lower redshifts. They are also in agreement with other recent \jwst\ constraints derived by \citet{whitler23} represented by purple circles. We note that for galaxy candidates at $z \gtrsim 11$, the absence of constraints on the Balmer break tend to underestimate the stellar mass we derive from SED fitting. Furthermore, the green-shaded region and the red-dashed line are theoretical predictions at $z=9$ computed from hydrodynamical zoom simulations \citep{kannan23} and semi-analytical models \citep{yung19}. Models also predict a rapid evolution with redshift, in line with our results.

%%%%%%%%%%%%%%%%%%%%%%%%%%%%%%%%%%%%%%%%%%%
\subsubsection{UV Continuum slopes}
\label{sec:beta}

\begin{figure}
    \centering
    \includegraphics[width=\columnwidth]{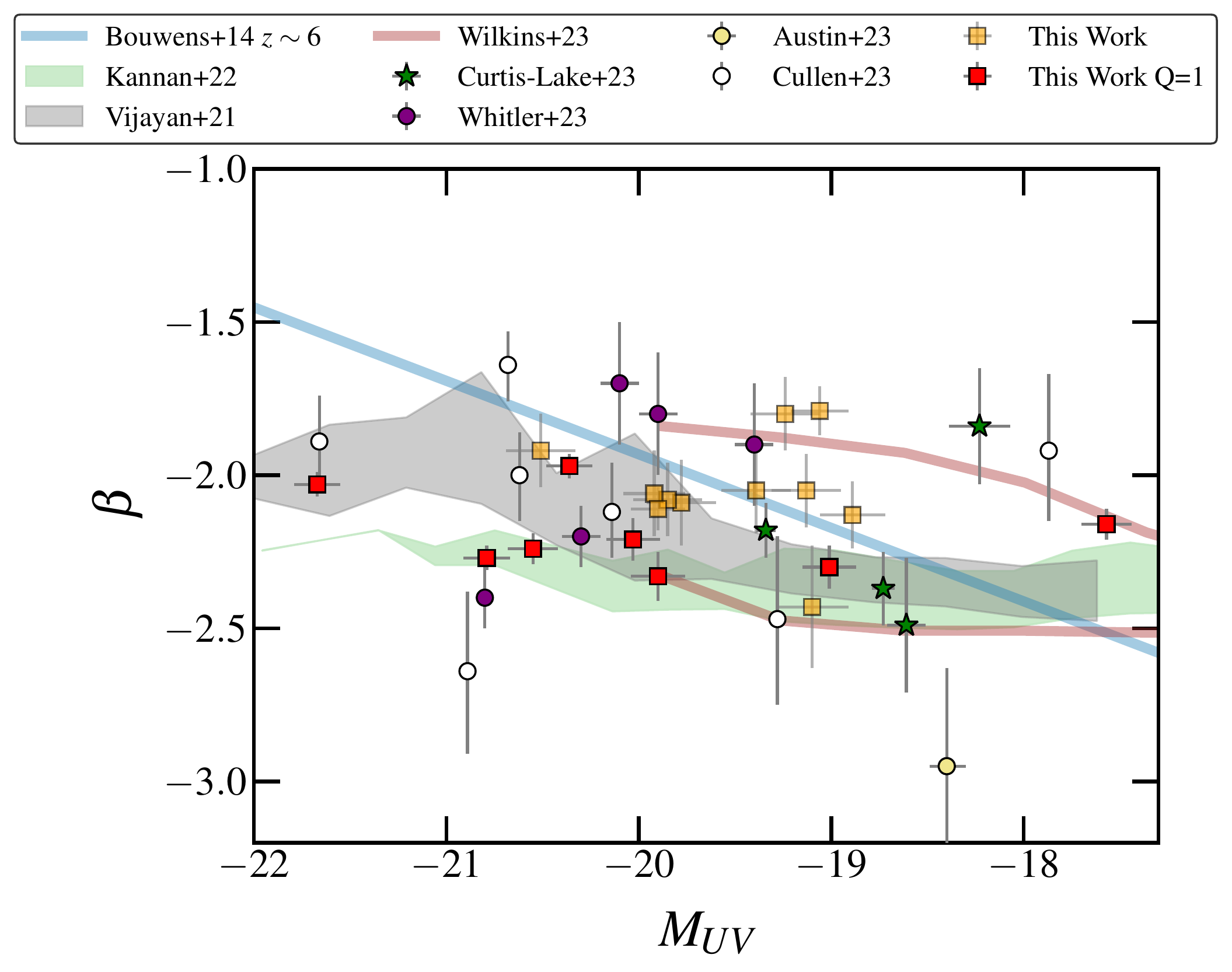}
    \caption{The UV continuum slope $\beta$ as a function of the UV magnitude. The present sample is represented by orange squares, while recent \jwst\ results at similar redshifts are marked with circles \citep{austin23,cullen23,whitler23} and stars \citep{curtis-lake23}. We also show theoretical predictions for this relation with a shaded green region \citep{kannan23} and a gray region \citep{vijyan21}, and two brown curves with (upper curve) and without (lower curve) dust attenuation \citep{wilkins23}. The empirical relation established at $z\sim6$ by \citet{bouwens14} is represented with the blue line.    
    }
    \label{fig:beta_muv}
\end{figure}

Next, we explore the UV continuum slope $\beta$, which is widely used to infer the dust attenuation in high-redshift galaxies. This parameter also encodes information about the age of the stellar population, where the contribution from young stars will result in a bluer UV slope. The $\beta$\ slope is measured by fitting a power law of the form $f_{\lambda} \propto \lambda^{\beta} $ to the rest-frame UV photometric measurements below 3000 \AA\ in F200W, F277W, F356W bands. We fixed the redshift to the {\tt Eazy} best-fit phot$-z$. To estimate the impact of redshift and photometric uncertainties, we performed a Monte Carlo sampling using a normal distribution around these fixed parameters in the fitting procedure. The results are provided in Table \ref{tab:sample}. Like most of the $z>9$ galaxies recently uncovered in \jwst\ observations, these candidates show blue UV slopes ranging from $\beta = -1.8$ to -2.3, which is expected if these candidates have younger stellar populations with low-durst attenuation and metallicities. Absence of dust has also been invoked to explain the high number density of high-z galaxies \citep{ferrara22}. We do not find however evidence for extremely blue slopes like those that were reported in early \jwst\ data \citep{topping22,adams23}. When compared to literature results (see Figure \ref{fig:beta_muv}), these candidates show bluer UV slopes at a given luminosity than those observed in $z \sim 6$ galaxies, for instance \citep{bouwens14} or more recently from \jwst\ observations \citep{nanayakkara23}.  They follow the general trend of $z>9$ galaxies \citep{austin23,cullen23,curtis-lake23,whitler23}, for which we show only robust measurements with uncertainties below 0.3 dex. Similarly, the $\beta$-\muv\ relation derived from the latest numerical simulations, such {\tt FLARES} \citep{vijyan21,wilkins23} and {\tt Thesan} \citep{kannan23}, show broad agreement with the observed $\beta$ values.  
 
In addition to the dependence with age, the UV slope is also affected by the nebular continuum, whose contribution is larger at longer wavelengths, which results in redder $\beta$ values. Albeit with large uncertainties, these effects, combined with constraints from nebular recombination lines, can be used to indirectly infer the escape fraction of ionizing continuum \fesc\ in galaxies at the epoch of reionization \citep[e.g.,][]{zackrisson17,plat19,topping22}.

%%%%%%%%%%%%%%%%%%%%%%%%%%%%%%%%%%%%%%%%%%%%%%%%%%%%%%%%%%%%%%%%%%%%%%
\section{Summary} 
\label{sec:conclusions}

In this paper, we presented the results of our search for $z>9$ galaxy candidates in the \jwst\ UNCOVER survey. We used deep NIRCam and NIRISS imaging from three observing programs UNCOVER \citep{bezanson22}, GLASS \citep{treu22}, and a DD program ID 2767, in addition to ancillary \hst\ observations. We combined dropout selection and photometric redshift estimates from two independent codes, {\tt Eazy} and {\tt BEAGLE}, to identify high-redshift galaxy candidates. We report the detection of 16 candidates at $9<z<12$ and 3 candidates at $12<z<13$. According to our quality assessment, a total of 7 candidates are deemed robust among this sample. Candidates span a wide dynamic range in luminosity, from \muv\ $\sim$ -22 to -17.6 mag. Some of these sources are among the faintest galaxies discovered at $z>10$, although most of their magnification factors are still relatively modest, i.e. below $\mu =5$. Two candidates have spectroscopic confirmation at redshift $z=9.76$ and $z=9.3$. 

In addition to photometric redshift, we ran refined SED fitting with {\tt BEAGLE} to constrain the physical properties of the sample, fixing the redshift and focusing on four parameters: the stellar mass, the stellar age, the star formation rate averaged over the last 10 Myr, and the attenuation $\tau_{V}$. We find that galaxies have young ages between 10 and 100 Myr and low star formation rates from $\sim 0.2$ to $\sim 7$ \msolyr. These results confirm previous findings in early \jwst\ observations of $z>9$ galaxies. Most of the candidates have low stellar masses in the range log(\mstar/\msol) $\sim$ 6.8-9.5. We find evidence for a rapid redshift-evolution of the mass-luminosity relation, in line with recent observational results and theoretical predictions.  

We also find that these galaxies have blue UV continuum slopes, between $\beta =-1.8$ and $\beta=-2.4$, although we do not find extremely blue $\beta$ values as measured in recent $z>10$ studies. The young ages we measure are consistent with these blue continuum slopes. We also see a redshift-evolution of the UV slope for a given intrinsic magnitude. The sample aligns with theoretical predictions and similar observational results for the $\beta$-\muv\ relation at $z>9$.   

In the near future, \jwst\ will continue to provide increasingly larger samples of rest-optical observations of galaxies at $z>9$. Combined with follow-up spectroscopy, these data will help constrain more precisely the physical properties and star formation histories of these galaxies. In particular, ultra-deep observations of additional lensing clusters will help us push to fainter and more representative galaxies at those early epochs.

\section*{Acknowledgements}

\noindent This work is based on observations obtained with the NASA/ESA/CSA \textit{JWST} and the NASA/ESA \textit{Hubble Space Telescope} (HST), retrieved from the \texttt{Mikulski Archive for Space Telescopes} (\texttt{MAST}) at the \textit{Space Telescope Science Institute} (STScI). STScI is operated by the Association of Universities for Research in Astronomy, Inc. under NASA contract NAS 5-26555. This work has made use of the \texttt{CANDIDE} Cluster at the \textit{Institut d'Astrophysique de Paris} (IAP), made possible by grants from the PNCG and the region of Île de France through the program DIM-ACAV+. This work was supported by CNES, focused on the \jwst\ mission. This work was supported by the Programme National Cosmology and Galaxies (PNCG) of CNRS/INSU with INP and IN2P3, co-funded by CEA and CNES. P. Dayal acknowledges support from the NWO grant 016.VIDI.189.162 (``ODIN") and the European Commission's and University of Groningen's CO-FUND Rosalind Franklin program.

\section*{Data Availability}
The data underlying this article are publicly available on the \texttt{Mikulski Archive for Space Telescopes}\footnote{\url{https://archive.stsci.edu/}} (\texttt{MAST}), under program ID 2561. Reduced and calibrated mosaics are also available on the UNCOVER webpage: \url{https://jwst-uncover.github.io/} 

\bibliographystyle{mnras}
\bibliography{references}

\end{document}